\documentstyle[12pt]{article}
\newcommand\mref[1]{(\ref{#1})}
\newcommand{\vol}{\mbox{Vol}M^3}
\newcommand{\etal}{{\it et al.\/}}
\newtheorem{theorem}{Theorem}
\newtheorem{lemma}{Lemma}

\newcommand\bC{{\bf C}}
\newcommand\bR{{\bf R}} 
\newcommand\bZ{{\bf Z}}

\begin{document}

\title{On the Integrability of Bianchi Cosmological Models}

\author{Andrzej J.~Maciejewski \\
Institute of Astronomy, 
Nicolaus Copernicus University \\
87-100 Toru\'n, Chopina 12/18, Poland\\
E-mail: maciejka@astri.uni.torun.pl 
\and
Marek Szyd{\l}owski \\
Astronomical Observatory, Jagiellonian University, \\
Orla 171, 30-244 Krakow, Poland \\
E-mail: szydlo@oa.uj.edu.pl}

\date{}

\maketitle

PACS number: 98.80

\begin{abstract}
In this work, we are investigating the problem of integrability of 
Bianchi class A cosmological models. This 
class of systems is reduced to the form of Hamiltonian systems 
with exponential potential forms. 

The dynamics of Bianchi class A models is investigated through 
the Euler-Lagrange equations and geodesic equations in the Jacobi 
metric. On this basis,  we have come to some general conclusions 
concerning  the evolution of the volume function of 3-space of 
constant time. The formal and general form of this function has been 
found. It can serve as a controller during numerical calculations 
of the dynamics of cosmological models.

The integrability of cosmological models is also 
discussed from the points of view of different integrability criterions.
We show that dimension of phase space of   Bianchi class A Hamiltonian systems
can be reduced by two. We prove vector field of the reduced system is polynomial and it does not admit any analytic, or even formal first integral. 
\end{abstract}

\newpage

\section{Introduction}

We shall investigate the dynamics of the most interesting group of 
homogeneous Bianchi class A cosmological models which is described 
by the natural Lagrangean function
\begin{eqnarray}
{\cal L} &=& \frac{1}{2} g_{\alpha \beta} 
\dot{q}^{\alpha} \dot{q}^{\beta} - V(q) = 
T - V(q)=\nonumber \\ 
\label{e:b9a}
& =&\frac{1}{4} \sum_{i=1,i<j}^{3} 
\frac{d \ln q^i}{dt} \frac{d \ln q^j}{dt} - 
\frac{1}{4} \left( 
2 \sum_{i=1,i<j}^{3} n_i n_j q^i q^j - 
\sum_{i=1}^{3} n_{i}^{2} q_{i}^{2} \right) ,
\end{eqnarray}
where 
\(
q_i \approx A_{i}^{2} (i=1,2,3)
\)
are three squared scale factors $A_i$ for diagonal class A Bianchi 
models; different Bianchi types correspond to different choices 
of $n_i\in\{ -1,0,1\}$, $i=1,2,3$;
a dot denotes differentiation with respect to cosmological time $t$. 
Logarithmic time $\tau$ is related with cosmological time $t$ by 
\[
d\tau = \frac{dt}{(q^{1} q^{2} q^{3})^{1/6}} 
= \frac{dt}{\mbox{Vol} M^{3}} \ \ .
\]

Bogoyavlensky \cite{kn:BogoyavlenskyN}, proves an important 
property of  system \mref{e:b9a}, namely the existence of 
the monotonic function $F$ with the following form
\begin{equation}
\label{e:F}
F = \frac{d}{dt} (q^{1} q^{2} q^{3})^{1/6} 
= \frac{d}{dt} \mbox{Vol}M^{3},
\end{equation}
such that 
\[
\frac{dF}{dt} \le 0.
\]
Function \mref{e:F} is invariant with respect to the scaling 
transformations and it has the sense of the speed of change 
of the average radius of the universe. Function $| F |$ 
along any solution decreases from infinity to zero in  such  a way that 
$F=0$ is reached at the moment of maximal expansion, and 
$| F | = \infty$ corresponds to the initial singularity. 
The existence of function $F$ allows us to define what we 
call the early stage of the evolution of the universe as
\[
F \gg 1 \ \ .
\]

The importance of this function for numerical integration of 
B(IX) models has been pointed out in the work \cite{kn:RughJ}. 
The authors have used the Rauchaudhuri equations to show 
the property of upper-convexity of function 
$(\vol)(t)$ which means that this function does not 
possess a local minimum (where $F=0$ and $\dot{F}>0$) and 
may possess not more than one maximum. If $F<0$, 
the volume function $(\vol)(t)$ shrinks; whereas 
if $F>0$ it expands. Both of the processes take place in the 
same region of the phase space $(p,q)$ but with reverse 
directions of time. In the phase space $(p,q)$ function 
$F$ has the following form
\[
F = \frac{(q^1 q^2 q^3)^{1/6}}{3} \sum_{i=1}^{3} p_i q^i, \qquad 
p_i = \frac{\partial L}{\partial \dot{q}^i},
\]
and
\[
\frac{dF}{dt} = \frac{(q^1 q^2 q^3)^{1/6}}{9} 
\left[   \left( \sum_{i=1}^{3} p_i q^i \right)^2 -6V\right],
\]
where $p_i$ 
are momenta conjugated with generalized coordinates $q^i$.

In  work \cite{kn:RughJ},  function $F$ 
was used for  controlling  the quality  
of numerical integrations of the B(IX) model. In this model, the scale 
factors  oscillate in a neighborhood 
of the initial and final singularity. The function 
$(\vol)(t)$,obviously, does not possess the 
analogous property \cite{kn:RughJ,kn:Rugh}.

In the present work we give some general and formal 
expressions for the function $(\vol)(t)$. It can be used  
 as a tool for studying the B(IX) models. Let us note 
recent important results of  Cushman and \'Sniatycki \cite{kn:CS} 
concerning chaos in the B(IX) system. They proved that existence of a monotonic 
function $F$ excludes the  possibility of recurrence  in the system and, thus, any form of standard deterministic chaos in the system. This illuminates previous negative results and shows  that for a study of this system 
we have to use non-conventional methods. 

Several authors tested   if the last model passes the standard Painlev\'e integrability test (in the form of the ARS algorithm \cite{kn:Ramani}). 
First results of Contopoulos \cite{kn:Contopoulos} shows that
B(IX) model passes this test.  Next, this paper 
was revised \cite{CGR1994}, however,  without any strict conclusions
concerning integrability.   It  was stated also that
 this model passes Ziglin's test (see \cite{Ziglin:82::a,Ziglin:82::b}).  More careful Painlev\'e analysis was done by 
Latifi \etal\  in \cite{kn:Conte}. They show that  B(IX) model does not passes 
the so called perturbative Painlev\'e test. Authors of this paper  suggest
the existence of `some chaotic r\'egimes' in the system.

The above remarks show that the notion `chaos' has  unclear status  when dynamical systems arising from  the general relativity and cosmology  are studied.   Moreover, for B(IX) model discrete dynamics defined in \cite{kn:Chernoff}
shows strong ergodic properties, however, this `chaotic behaviour' seems to 
be absent (or hidden) in  the continuous dynamics. 
Moreover, the standard criteria of detection of chaos (Lyapunov 
characteristic exponents---LCE) are not invariant with respect 
to the time reparametrization and transformation of phase 
variables whereas existence of first integrals is an invariant 
property of the system.  It is also important to note that the 
non-zero LCE can be used as an indicator of chaos only when the 
motion take place in a compact invariant subset of the phase space,
 but it is not true for the B(IX)
dynamical system. All  these facts   motivated us  to study the problem of
integrability of the investigated models.

The non-integrability of the system is a weaker property than 
chaos (in the sense of the deterministic chaos) but better described 
and understood. The authors believe that 
investigation of non-integrability in B(IX) models can contribute 
to a better understanding of chaos in cosmological models. 
Here we show that the Bianchi class A Hamiltonian system are not completely integrable in the sense of Birkhoff. This  conclusion is weak as the 
negative answer to the question about algebraic  complete integrability 
of B(IX) (see \cite{Pavlov:96::}). In order to obtain stronger result we 
reduce the dimension of the phase space by two. We show that the 
reduced system is polynomial and, what is most important, it does not 
admit any analytical, or even formal, first integral. 

In cosmological  models chaos, if properly defined and  present, 
has some hidden character.
The basic indicator of chaos in these models, the LCE,  depends on the choice of the time parametrization. In 
the logarithmic time $\tau$,  nearby trajectories  diverge linearly
 whereas in other  time parametrizations 
they will  diverge  exponentially which is characteristic for 
 chaotic systems. The fact that the rates of separation of nearby 
trajectories depend on the clock used is obvious. 
The problem is in invariant choices of the time parameter 
for the invariant chaos detection. Such a role is played 
by Maupertuis clock (time parameter $s$ is such that 
$\frac{ds}{d\tau} = 2|E-V|$, where $E$ is the total energy 
of the system, $V$ is its potential and $\tau$ is 
mechanical time).

Our point of view is such that the
LCE, when used in  general relativity, 
should be defined in an invariant way. Then the results 
could be interpreted in a different time parametrizations. 
The Bianchi IX model is `chaotic' in the parameter $s$ 
(LCE is positive), but, after transition to the parameter $\tau$,  
nearby trajectories diverge linearly in such a way as integrable 
systems. This phenomenon is called the hidden chaos. 
Let us note that the existence of the first integral of 
an autonomous system is an invariant property (with respect to 
time reparametrization and to transformation of phase variables).

In general relativity and cosmology, the problem of 
non-integrability or chaos is not only very subtle but also is  strictly connected 
with the invariant description. One must be very careful detecting 
integrability in B(IX) dynamics.  The problem 
whether chaos in the gauge theory is a physical phenomenon 
is, generally,  open.

\section{The dynamics of Bianchi class A models from 
the Euler-Lagrange equations}

The Hamiltonian function for the system (1) has 
the following form
\begin{equation}
\label{e:H}
{\cal H} = \frac{1}{2} g^{\alpha \beta} 
p_{\alpha} p_{\beta} + V(q),
\end{equation}
where
\[
p_{\alpha} = g_{\alpha \beta} \dot{q}^{\beta} ,
\]
\[
g^{\alpha \beta} = 2 
\left( \begin{array}{lll} 
-(q^1)^2 & q^1 q^2 & q^1 q^3 \\
q^2 q^1 & -(q^2)^2 & q^2 q^3 \\
q^3 q^1 & q^3 q^2 & -(q^3)^2
\end{array} \right),
\]
\[
V(q) = \frac{1}{4} \left( 2 \sum_{i<j}^{3} 
n_i n_j q^i q^j - \sum_{i=1}^{3} n_{i}^{2} 
(q^i)^2 \right),
\]
$V(q)$ is the potential function. The obtained Hamiltonian system  is considered  
only on an invariant set of the phase space defined by
the  zero level of the Hamiltonian \mref{e:H}, i.e.,
\begin{equation}
\label{e:H0}
{\cal H} = 0.
\end{equation}
The Euler-Lagrange equations in time $\tau$ have the following 
form
\begin{equation}
\label{e:E-L}
\frac{d^2 q^{\alpha}}{d\tau^2} + \Gamma^{\alpha}_{\beta \gamma} 
\frac{dq^{\beta}}{d\tau} \frac{dq^{\gamma}}{d\tau} = 
g^{\alpha \beta} \frac{\partial V}{\partial q^{\beta}},
\end{equation}
where Christoffel symbols \( \Gamma^{\alpha}_{\beta \gamma} \) 
are connected with the metric defined by the kinetic energy 
\[
T = \frac{1}{2} g_{\alpha \beta} \dot{q}^{\alpha} \dot{q}^{\beta} 
= \frac{1}{2} g^{\alpha \beta} p_{\alpha} p_{\beta}.
\]
Equations \mref{e:E-L} after transformations to a new time 
parameter $s$, called Maupertuis time, take the form 
of geodesic equations for  the Jacobi metric 
\[
\hat{g}_{\alpha \beta} = 2 | E-V | g_{\alpha \beta} 
= 2 W g_{\alpha \beta}, 
\]
that is, 
\begin{equation}
\label{e:sE-L}
\frac{d^2 q^{\alpha}}{ds^2} + \hat{\Gamma}^{\alpha}_{\beta \gamma} 
\frac{dq^{\beta}}{ds} \frac{dq^{\gamma}}{ds} = 0,
\end{equation}
where a hat denotes that respective quantities are calculated 
with respect to the Jacobi metric. Christoffel symbols 
calculated from  $g$ and $\hat{g}$ metrics are connected
by relations

\begin{equation}
\hat{\Gamma}^{i}_{jk} = \Gamma^{i}_{jk} + A^{i}_{jk}, 
\end{equation}
where
\[
A^{i}_{jk} = (\partial_j \Phi) \delta_{k}^{i} + 
(\partial_k \Phi) \delta_{j}^{i} - 
g^{ir} (\partial_r \Phi) g_{jk} ,
\]
\[
\Phi = \frac{1}{2} \ln 2W.
\]
Let us note that the kinetic energy form does not depend on 
the Bianchi type models characterized by the set $\{n_1,n_2,n_3\}$. The only 
non-vanishing Christoffel symbols are
\begin{equation}
\label{e:nz}
\Gamma^{1}_{11} = - \frac{1}{q^1} , \qquad
\Gamma^{2}_{22} = - \frac{1}{q^2} , \qquad 
\Gamma^{3}_{33} = - \frac{1}{q^3} .
\end{equation}
After substitution \mref{e:nz},  system \mref{e:E-L} takes the form 
\begin{equation}
\frac{1}{q^i} \frac{d^2 q^i}{d\tau^2} - 
\left(\frac{d}{d\tau} \ln q^i \right)^2 = 
(n_j)^2(q^j)^2 + (n_k)^2(q^k)^2 - (n_i)^2(q^i)^2 
- 2 n_j n_k q^j q^k,
\end{equation}
where $\{i,j,k\}\in S_3$, and   $S_3$ denotes the set of even  permutations of $\{ 1,2,3\}$.  

The change of variables
\begin{equation}
q^i = e^{Q^i},\qquad i =1,2,3,
\end{equation}
transforms  the above equations to the following form
\begin{equation}
\label{e:Q}
\frac{d^2 Q^i}{d\tau^2} = 
(n_j)^2 e^{2Q^j} + (n_k)^2 e^{2Q^k} - (n_i)^2 e^{2Q^i} 
- 2 n_j n_k e^{Q^j + Q^k}, 
\end{equation}
where  $\{i,j,k\}\in S_3$.

The system \mref{e:Q} is satisfied on the Hamiltonian constraint 
${\cal H} = 0$ which is equivalent to the condition 
of normalization of the tangent vector to the trajectory 
\( u^i = dq^i/ds \), that is,
\begin{equation}
\| u \|^2 = 2W g_{\alpha \beta} 
\frac{dq^{\alpha}}{ds} \frac{dq^{\beta}}{ds} = 
- \mbox{sgn} V,
\end{equation}
or
\[
g_{\alpha \beta} 
\frac{dq^{\alpha}}{d\tau} \frac{dq^{\beta}}{d\tau} = 
2V \mbox{sgn} V.
\]
In terms of variables $Q^i$, the constraint 
condition is equivalent to
\begin{equation}
\sum_{i<j}^{3} 
\frac{dQ^i}{d\tau} \frac{dQ^j}{d\tau} = - 8V.
\end{equation}
Adding the sides of equations \mref{e:Q}, we obtain 
the following formula
\begin{equation}
\label{e:sumQ}
\sum_{i=1}^{3} 
\frac{d^2Q^i}{d\tau^2} = - 4V.
\end{equation}
Equations \mref{e:Q}, after time reparametrization 
\( \tau \rightarrow s = s(\tau) \),
take the form of geodesic equations. From \mref{e:Q} we obtain 
\[
4W^2 \frac{d^2 Q^i}{ds^2} + \frac{dQ^i}{ds} 
\frac{\partial W}{\partial Q^j} \frac{d Q^j}{ds} 
= (n_j)^2 e^{2Q^j} + (n_k)^2 e^{2Q^k} - (n_i)^2 e^{2Q^i} 
- 2 n_j n_k e^{Q^j + Q^k},
\]
where $\{i,j,k\}\in S_3$.

The problem of investigation of Lagrange systems with 
the indefinite kinetic energy form is open. 
First steps to investigate such systems have been 
done in \cite{kn:SzydloSH}. In the terminology used in 
\cite{kn:SzydloSH} our system is a special case, the so called 
 non-classical simple mechanical system. As it was established, 
these systems have the following fundamental  property. A trajectory 
of the system can pass through the set $\partial D =\{ q: E-V=0 \}$. During this passage the  vector tangent to 
 trajectory  changes   the cone sector defined  by the kinetic 
energy form: 
\( g_{\alpha \beta} (q_0) \xi^{\alpha} \xi^{\beta} = 0 \) 
where \( \xi^{\alpha} = dq^{\alpha} /ds \), 
\( q_0 \in \partial D$. In our case the signature of $g_{\alpha \beta}$ 
is Lorentzian, i.e., $(-,+,\ldots,+)$ 
(for details see \cite{kn:SzydloSH}). 

In  generic situations 
($n_i \ne 0$ for $i=1,2,3$) which include BVIII and BIX models (Mixmaster 
models), there are analytical and numerical arguments  that the function of sign of the potential for a typical trajectory is an infinite subsequence
that is a one sided cut of the following   double infinite 
sequence (see \cite{kn:Burd})
\begin{equation}
\label{eq:seq}
\mbox{sgn} V = \{ \ldots, +1, 0, -1, 0, -1, \ldots \}.
\end{equation}
If we assume that the subsequence \mref{eq:seq} is finite, then our 
system reaches the state $V=0$ ($W=0$ in general case) 
which corresponds to Kasner solutions, finite number of 
times. During the Kasner epoch the information about 
the localization of the point on the interval of normal 
separation (modulo initial localization) grows $e$-times 
\cite{kn:Bogoyavlensky}. If the subsequence \mref{eq:seq} is finite 
it means that after $\bar{n}$ epochs $\bar{n}$ bytes (i.e., 
finite number of bytes) of information have been lost 
whereas we know that our system is chaotic (the loss of 
infinite information is required for chaos). Let us notice 
that when the system goes asymptotically to the boundary sets 
$W=0$ then it is asymptotically free.

\section{Properties of the function of volume of the   constant time 3-space }

Equation  \mref{e:sumQ} implies that
\begin{equation}
\frac{1}{2} \frac{d^2}{d\tau^2} 
\ln (\mbox{Vol}M^3) = -2V = 2T .
\end{equation}
The above relation means that, in a generic case $(\forall i\, n_i \ne 0)$,  there is an infinite 
number of intervals in which function $\ln (\mbox{Vol}M^3)$ is subsequently convex up and 
down. These intervals are separated by an  infinite number of 
inflexion points (which corresponds to $V=0$) in the diagrams of 
the function $\ln (\mbox{Vol}M^3)(\tau)$ and lie on the lines 
$\ln (\mbox{Vol}M^3)(\tau) = \pm \tau + C$. The additional 
information we have about the B(IX) model  
is that this model has the initial and final singularities. In the following 
paragraphs of the work we shall concentrate on the B(IX) models.

From the $(0,0)$ components of the Einstein equation for the 
BIX case, we obtain that  function $\ln (\mbox{Vol}M^3)(\tau)$ 
cannot possess a local minimum but it can possess a single 
maximum. The above property suggests that for a  typical trajectory 
in this model the qualitative diagrams of the function 
$\ln (\mbox{Vol}M^3)(\tau)$ look like in Figure 1.

Fig. 1.

After integrating over $\tau$ the both sides of \mref{e:sumQ} and 
assuming that in the moment of maximal expansion $\tau = 
\tau_0$, we obtain
\begin{equation}
\label{e:defln}
\ln (\mbox{Vol}M^3)(\tau) \propto e^{C_1 \tau} 
e^{\int_{\tau_0}^{\tau} s(t) \mbox{sgn}(-V) dt},
\end{equation}
where we choose $C_1 =1$ in the phase of expansion and 
$C_1 = -1$ in the phase in contraction of the volume function 
(if $V=0$, $\ln (\mbox{Vol}M^3)(\tau) \propto e^{\pm \tau}$).
Finally, for any model which describes the evolution of the 
volume function
\begin{equation}
\label{e:ln}
\ln (\mbox{Vol}M^3)(\tau) \propto e^{\tau (1 + \langle s \rangle)} 
\rightarrow_{\tau \rightarrow \infty} e^{\tau} e^{\langle s \rangle} 
\propto e^{\tau},
\end{equation}
\[
\langle s \rangle = \frac{1}{\tau} \int_{\tau_0}^{\tau} 
s(t) \mbox{sgn} (-V) dt,
\]
where we assume that the average value of $s(\tau)$ on the interval 
$(\tau, \tau_0)$ exists as $\tau \rightarrow - \infty$,  and it is 
finite. From the formula \mref{e:ln} it immediately yields that in a 
neighborhood of the initial singularity $(\tau \rightarrow 
- \infty)$ the volume function changes exactly as in Kasner's models. 
The second observation is as follows: the volume function 
does not oscillate around the equilibrium positions 
$\ln (\mbox{Vol}M^3)(\tau) \equiv 0$ but oscillates around 
Kasner's solution. In other words, Kasner's solution plays 
the role analogical to the equilibrium positions in the small 
oscillation approximation.

From \mref{e:defln} one can obtain the following relations between a 
natural parameter $s$ defined along geodesics (Maupertuis time) and the 
volume function $\ln (\mbox{Vol}M^3)(\tau)$
\begin{eqnarray*}
s & = \displaystyle\frac{d}{dt} \ln (\mbox{Vol}M^3)(\tau) & \qquad
\mbox{for} \qquad V<0 , \\
s & = - \displaystyle\frac{d}{dt} \ln (\mbox{Vol}M^3)(\tau) & \qquad
\mbox{for} \qquad V>0.
\end{eqnarray*}
The zero value of the parameter $s$ corresponds to  
the moment $\tau = \tau_0$. The relations between the parameter $s$,  
function $F$ and the scalar expansion function 
\[
 \Theta \equiv \frac{d}{dt} \ln (\mbox{Vol}M^3)(t), 
\]
 are as  follows 
\begin{eqnarray*}
s(\tau) & = & \pm 6 [(\mbox{Vol}M^3)(\tau)]^{1/3} F(\tau), \\ 
s(\tau) & = & \pm  [(\mbox{Vol}M^3)(\tau)] \Theta(t(\tau)), 
\end{eqnarray*}
where plus and minus sign correspond to  $V<0$, and $V>0$,
respectively.

From the above, we can conclude that "near the singularity" is 
equivalent to $s \gg (\mbox{Vol}M^3)^{1/3}$, i.e., $s \gg 0$.

Formula \mref{e:ln} implies that the characteristic time after 
which $(\mbox{Vol}M^3)(\tau)$ grows $e$-times, i.e., 
\( (\mbox{Vol}M^3)(\tau) \propto e^{\tau/\tau_{char}} \) 
has the following form
\begin{equation}
\tau_{char} = (1 + \langle s \rangle )^{-1} .
\end{equation}
This characteristic time is finite if the average $\langle s \rangle$ 
exists. 

For a typical trajectory  the function 
$s(\tau)$ is shown in Fig. 2.

\section{The Bianchi class A models as systems with exponential 
potentials and Its Algebraic Non-integrability}

After introducing the new variables $Q^i$ and using the definition 
(10), the Lagrange system (1) can be transformed to a Hamiltonian 
one. The Hamilton function for this system takes the following form
\begin{eqnarray}
{\cal H}(p, Q) & = & 2 \sum_{i<j}^{3} p_i p_j - 
\sum_{i=j}^{3} p_{i}^2 + \frac{1}{4} 
\left( 2 \sum_{i<j}^{3} n_i n_j e^{Q^i + Q^j} - 
\sum_{i=j}^{3} n_{i}^{2} e^{2Q^i} \right) \nonumber \\
\label{e:eham}
& = & \frac{1}{2} g^{\alpha \beta} p_{\alpha} p_{\beta} + 
V(Q^{\alpha}) ,
\end{eqnarray}
where $p_i = \frac{1}{4}(\dot{Q}^j + \dot{Q}^k)$ for  $\{i,j,k\}\in S_3$.
The Hamilton function \mref{e:eham} is a special case of the Hamiltonian 
for the  so called perturbed Toda lattice \cite{kn:Bogoyavlensky}. 

\section{Analysis of the integrability of B(IX) model}

There are several definitions of integrability. Generally, integrability means 
that the system under consideration possesses a large enough number of first 
integrals. Necessarily we have to specify the class of functions that contains  these 
first integrals as well as to define  the domain of their definition. 
Let us note here  that there are   examples of Hamiltonian systems that possess
an first integral of class $C^\alpha$ but does not possess an integral of class 
$C^{\beta}$ with $\beta>\alpha$, for $\alpha, \beta = 1, 2, \ldots \infty, \omega$ (see \cite{Kozlov:87::b}). 

It is well known also that every system of $n$ differential autonomous equations  is locally integrable---in a neighborhood of 
every nonsingular point (where the right hand sides do not vanish) it possesses $n-1$ first integrals. Thus, nontrivial problems are non-local or concern the  existence of integrals in a neighborhood of equilibria points.

It is very difficult to prove (non)integrability of a given set of differential 
equations. One way to simplify the problem is to restrict a class of function where we look for integrals.    As an illustration 
of this  approach, let us consider the Birkhoff integrability (see \cite{Kozlov:88::}) of the Bianchi class A  system in the form \mref{e:eham}.
This system belongs to the wide class of Hamiltonian systems in $\bR^{2n}$ equipped with the  standard symplectic structure and are given by the following  Hamiltonian function 
\begin{equation}
\label{e:spec}
  {\cal H} = \frac{1}{2} ( p, p)  + \sum_{m\in {\cal M}} v_m \exp\langle c_m,q\rangle ,
\end{equation}
where
\[
( p, p) = \sum_{i,j=1}^na^{ij}p_i p_j, \qquad \langle c_m,q\rangle= \sum_{i=1}^nc_{m_i}q^i, \quad m =(m_1, \ldots, m_n)\in \bZ^n,
\]
$(a^{ij})$, $v_m$ and $c_m$ are constant;  ${\cal M}$ is a finite subset of $\bZ^n$ 
\[
 {\cal M} =\{ m\in\bZ^n\; | \;  v_m\neq 0 \}.
\]
We look for integrals that are polynomials with respect to $p$, i.e., 
\[
     f(q,p) = \sum_{k\in {\cal N}_l}f_k p^k,
\]
where 
\[
{\cal N}_l =\{ m\in\bZ_+^n\; | \; |m|\leq l\}, \qquad |m|= \sum_{i=1}^nm_i; \qquad p^m = p_1^{m_1}\ldots p_n^{m_n }, \quad m \in \bZ_+^n,
\]
and coefficients $f_k$ have the form of infinite series of exponents
\[
f_k= \sum_{m\in \bZ^n} f^{(k)}_m \exp (c^{(k)}_m,q).
\]
Here $\bZ_+$ denotes non-negative integers. We say that the system \mref{e:spec} is Birkhoff integrable if it possesses 
$n$ independent integrals of the prescribed  form (see note of Ziglin \cite{Ziglin:91::} about modification of the original definition of Kozlov). 
We  order elements of ${\cal M}$ with respect to the lexicogrphic order and 
denote by $\alpha$ its maximal element and by $\beta$ the maximal element of $M$ that is not colinearly with $\alpha$. Then, according to Theorem 3 from \cite{Kozlov:89::c} if 
\begin{equation}
\label{e:cond}
 k(\alpha,\alpha) + (\alpha,\beta)\neq 0 , \qquad \mbox{for all}\quad{k\in \bZ_+}
\end{equation}
then the Hamiltonian system \mref{e:spec} is not integrable in the sense of 
Birkhoff. 
We immediately have 
\begin{theorem}
A generic case  of the Bianchi class A  system given by Hamiltonian function \mref{e:eham} with  $n_i\neq 0$ for $i=1,2,3$  is not integrable in the sense 
of Birkhoff.
\end{theorem}
{\em Proof.} For the Hamiltonian  \mref{e:eham} we have 
\[
      {\cal  M}  = \{ (1,1,0), (1,0,1), (1,0,0), (0,1,1), (0,1,0), (0,0,1) \},
\]
and thus $\alpha = (1,1,0)$  and $\beta = (1,0,1)$. Metric $
(a^{ij})$ has the form 
\[
 (a^{ij}) = 2 \left[\begin{array}{rrr}
                  -1 &  1 & 1\\
                   1 & -1 & 1 \\
                   1 &  1 & -1
                  \end{array}\right],
\]
and thus we have
\[
k(\alpha,\alpha) + (\alpha,\beta)= 4 \neq 0,
\]
and this finishes the proof. $\Box$

Let us remark that in a case when one of $n_i$ is equal zero then $k(\alpha,\alpha) + (\alpha,\beta)=0$ for all $k$. In such a case the system 
has one additional integral, namely $p_i$.  

\section{ Reduction and Non-integrability of B(IX) model}
The results obtained in the previous section are weak. There are two reason 
of it. First, we asked about the {\em complete} integrability of the system. 
However, the system under investigation can have  only {\em one additional integral}. The most important is the fact the we pose our question 
for the system defined on $\bR^6$ although  we are interesting only in 
the system on five dimensional manifold defined by the level ${\cal H}=0$.
One can imagine a system that is not globally integrable although it is integrable on a one prescribed energy surface. 

In this section  we  want to  study 
the B(IX) Hamiltonian system just only the level ${\cal H}=0$. 
 
During investigation of a dynamical system we usually try 
to  lower its dimension making use of its first integrals and symmetries.
For Hamiltonian system \mref{e:H} we know only one first integral---Hamiltonian.
Thus, using it  we can potentially can make reduce the dimension of the 
system by one however we loose  the polynomial form of the system.  However we 
do not want this side effect of reduction because it excludes  possibilities 
of applications algebraic tools for study non-integrability.

In this section we show how to reduce the dimension of the phase space 
by two and preserving the polynomial form of the considered vector field.
In what follows  consider  her the case of B(IX) ($n_1=n_2=n_3=1$).

First, we transform the Hamiltonian vector field corresponding to Hamiltonian 
\mref{e:H}, to a homogeneous polynomial form of degree two.  To this end 
let us put 
\begin{equation}
\label{}
y_i = q^i, \qquad z_i = \frac{\dot q^i}{q^i}, \qquad i=1,2,3,
\end{equation}
then equation of motion will have the form
\begin{equation}
\label{e:yz}
\dot y_i = y_iz_i, \qquad \dot z_i = (y_j - y_k)^2 - y_i^2, \qquad \{i,j,k\}\in S_3. 
\end{equation}
 This system has the first integral corresponding to the Hamiltonian \mref{e:H}.
It is has the form 
\begin{equation}
\label{e:hyz}
H =  z_1 z_2 + z_1 z_3 + z_2 z_3
  -y_1^2  + 2y_1 y_2 -  y_2^2  + 2y_1 y_3 + 2 y_2 y_3 - y_3^2.  
\end{equation}
We make change of variables
\begin{equation}
w_1 = y_1 + y_2, \quad w_2 = y_1 - y_2, \quad w_3 = y_3,  
\end{equation}
and we leave $z_i$ unchanged. In new variables system \mref{e:yz}
has the  form 
\begin{equation}
\left.
\begin{array}{l}
 \dot w_1 = \frac{1}{2}z_1(w_1 +w_2)+ \frac{1}{2}z_2(w_1 -w_2),\\[\bigskipamount]
 \dot w_2 = \frac{1}{2}z_1(w_1 +w_2)- \frac{1}{2}z_2(w_1 -w_2),\\[\bigskipamount]
 \dot w_3 = z_3w_3,\\[\bigskipamount]
 \dot z_1 = (w_3 - w_1)(w_2 + w_3),\\[\bigskipamount]
 \dot z_2 = (w_3 - w_2)(w_3 - w_1),\\ [\bigskipamount]
 \dot z_3 = (w_3 + w_2)(w_2 - w_3), 
\end{array}
\right\}
\end{equation}
and the first integral \mref{e:hyz} is transformed to the following form  
\begin{equation}
H =  z_1 z_2 + z_1 z_3 + z_2 z_3
  -w_2^2  + 2w_1 w_3 -  w_3^2.  
\end{equation}
Now, we introduce new variables 
\begin{equation}
u_1 = \frac{z_1}{w_3}, \quad u_2 = \frac{z_2}{w_3}, \quad u_3 = \frac{z_3}{w_3}, \quad u_4 = \frac{w_2}{w_3}, \quad u_5 = \frac{w_1}{w_3}, \quad u_6 = {w_3}. 
\end{equation}
After this transformation we obtain the following system 
\begin{equation}
\label{e:uu}
\left.
\begin{array}{l}
 \dot u_1 = u_6[ (1 + u_4)(1 - u_5)  -u_1 u_3 ] ,\\[\bigskipamount]
 \dot u_2 = u_6[ (1 - u_4)(1 - u_5)  -u_2 u_3 ] ,\\[\bigskipamount]
 \dot u_3 = u_6( u_4^2 -u_3^2 -1)\\[\bigskipamount]
 \dot u_4 = \frac{1}{2}u_6 [ u_4(u_1+u_2-2u_3)+u_5(u_1-u_2)]\\[\bigskipamount]
 \dot u_5 = \frac{1}{2}u_6 [ u_4(u_1-u_2)+u_5(u_1+u2-2u_3)u_5]\\[\bigskipamount]
 \dot u_6 = u_3u_6^2,
\end{array}
\right\}
\end{equation}
with the first integral 
\begin{equation}
\label{e:hu}
H = u_6^2 ( u_1 u_2 + u_1 u_3  + u_2 u_3 - u_4^2 + 2 u_5-1).
\end{equation}
Now, we make use the fact that B(IX) model is  considered only 
on the level $H=0$.  From equation $H=0$  we find $u_5$ as a function 
of $(u_1, u_2, u_3, u_4)$:
\[
u_5 = \frac{1}{2}( 1 + u_4^2 - u_1 u_2 - u_1 u_3  -u_2 u_3),
\]
thus we can eliminate this variable from the right hand sides of \mref{e:uu}. 
Moreover if we change the independent variable  according to the rule

\[
    \frac{d}{dt} = \frac{u_6}{2}\frac{d}{ds},
\]
(note that $u_6>0$) than the first four  equations  in \mref{e:uu} separate 
from the last two. Thus, we finally obtained the following close system describing the dynamic of the B(IX) model:
\begin{equation} 
\label{e:uuu}
\left.
\begin{array}{l}
 \dot u_1 =  (1 + u_4)[1 + u_1 u_2 +u_3( u_1   + u_2) -u_4^2]  -2u_1 u_3 ,\\[\bigskipamount]
 \dot u_2 =  (1 - u_4)[1 + u_1 u_2 +  u_3( u_1   + u_2) -u_4^2]  -2u_2 u_3  ,\\[\bigskipamount]
 \dot u_3 = 2( u_4^2-u_3^2 -1)\\[\bigskipamount]
 \dot u_4 =  u_4(u_1+u_2-2u_3)+\frac{1}{2}(u_1-u_2)[1-  u_1 u_2 - u_3( u_1  + u_2) + u_4^2 ].
\end{array}
\right\}
\end{equation}
This system will be called the reduced B(IX) system. We consider this system 
in $\bC^4$ 

\begin{theorem}
The reduced B(IX) system does not have a non-trivial analytic first integral. 
\end{theorem}

Our theorem will be a consequence of the following lemma.
\begin{lemma}
Consider a system of differential equation 
\begin{equation}
 \dot x = f(x), \quad f(0) = 0, \qquad f(x) = (f_1(x), ldots, f_n(x))\qquad x\in \bC^n,
\end{equation}
with analytic right hand side, with 
\begin{equation}
f(x) =  Ax + O(|x|^2) ,
\end{equation}
where matrix  $A$ has eigenvalues $\lambda_i\in\bC$, $i=1,\ldots, n$.  
If the system possesses an analytical first integral $F$ then there 
exist non-negative integers $i_1, \ldots, i_n$ such that
\begin{equation}
\label{e:r}
\sum_{k=1}^n i_k \lambda_k = 0, \qquad \sum_{k=1}^n i_k >0
\end{equation}
\end{lemma}
Let us assume that an analytic fist integral exist and that condition \mref{e:r} is not satisfied.  We represent  the first integral in the  following form
\[
  F = \sum_{l=k}^\infty F_l, \qquad F_k\neq 0, \quad k\geq 1,
\]
where  $F_l$ is a homogenous form of degree $l$
\begin{equation}
F_l =\sum_{i_1+ \cdots+i_n=l}F^{(l)}_{i_1,\ldots,i_n} x_1^{i_1}\cdots x_n^{i_n}, \qquad i_k \in \bZ_+,\quad k=1, ldots, n.
\end{equation}
From the equation 
\[
  \sum_{j=1}^n f_j(x) \partial_j F = 0, 
\]
we conclude that the form $F_k$ is a first integral of the system 
$\dot x = A x$, i.e.,
\begin{equation}
\label{e:l}
  \sum_{j=1}^n l_j(x) \partial_j F_k = 0, \qquad 
l_i(x) = \sum_{j=1}^n A_{ij}x_j
\end{equation}
If matrix $A$ is diagonalizable then we can assume that 
$ l_i(x) = \lambda_i x_i$, and  then equation \mref{e:l} reads
\begin{equation}
\sum_{i_1+ \cdots+i_n=k}F^{(k)}_{i_1,\ldots,i_n}\left[\sum_{l=1}^n i_l \lambda_l\right]  x_1^{i_1}\cdots x_n^{i_n}=0. 
\end{equation}
This equation implies that 
\[
F^{(k)}_{i_1,\ldots,i_n}\left[\sum_{l=1}^n i_l \lambda_l\right]= 0
\quad  \mbox{for all} \quad (i_1, \ldots, i_n)\in\bZ^n_+, \sum_{l=1}^n i_l =k,
\]
Because, $F_k\neq 0$ there exit indices $(i_1, \ldots, i_n)$ such that 
$F^{(k)}_{i_1,\ldots,i_n}\neq 0$ and that for such indices we 
have 
\[
\sum_{l=1}^n i_l \lambda_l=0.
\]
Contradiction with our assumption prove the Lemma for the case of a diagonalizable matrix $A$. In the case of non-diagonalizable matrix the prove 
is only technically more difficult. See \cite{Nowicki:94::}  and especially \cite{Furta:96::} where this approach is generalized. 

Let us remark the the above Lemma is also true if we assume the first integral 
is a formal power series. 

To prove our theorem let us notice that for the reduced B(IX) model  point $ z = (-i,-i,i,0)$ is an equilibrium point and the  matrix of the linearized system 
is diagonalizable and has  possesses  the  eigenvalues $(-2 i, -2 i, -4 i, -4 i)$. For this eigenvalues condition \mref{e:r} cannot be satisfied and imply 
that the system does not have an analytic first integral. In fact we prove more, namely, 
the system does not have a first integral that can be expanded around the point
$z$ into a formal power series.

\section{Conclusions}

The particular integrable subclasses of the Bianchi models play 
an important role in the analysis of the dynamics of the cosmological 
models. To illustrate this fact, let us consider the phase space of 
the solutions of Bianchi models in the Bogoyavlensky approach 
\cite{kn:BogoyavlenskyN}. In the Bogoyavlensky methods of 
investigations of the corresponding dynamical systems,  we glue 
to the phase-space the boundary $\Delta$ onto which 
the  system  prolongs almost 
everywhere. The systems on the boundary $\Delta$ can be integrated 
and, in this way, we can study basic properties  
of the trajectories near the singularity. From the existence of 
the monotonic function $F$, we obtain that in the generic situation 
($\forall i$ $n_i \ne 0$) the trajectories of the Bianchi class A 
models close up to the boundary $\Gamma$ as $F \ll -1$.

Now, the trajectories move along the corresponding ones lying 
on the boundary. All the trajectories are the separatrices of 
the critical points. At last the trajectories reach the 
neighborhood of  critical points $K$ (corresponding to 
the Kasner asymptotics of the space-time metric) and they begin 
to move along their separatrices. The corresponding space-time 
metric for the mixmaster models is the BKL approximation 
\cite{kn:Bogoyavlensky}. 

In this way the chaotic systems (so non-integrable) in 
Bogoyavlensky's approach can be well approximated by an 
 integrable system. This feature of so surprisingly 
good approximation  is not so far understood completely.

In Bianchi models with chaos we have the infinite series 
of Kasner epochs (it would be good if we had a precise 
and exact proof of this fact) and these models do not 
exactly admit the Kasner asymptotics.

It may be the case  when we consider the higher-dimensional 
generalization of the mixmaster cosmological models 
(or models with a massless scalar field). These systems 
admit exact Kasner asymptotics, so, according to 
our theorem they will be integrated.

In the work \cite{kn:Pereseckii} it has been proved 
that the BKL approximation is true not only for 
BVIII and BIX, but it exists for all Bianchi types 
(exceptionally BI and BV) with the movement of 
matter ($n^{\alpha} \ne \delta_{0}^{\alpha}$).

In the models I, V and also III, VII and the type II model, 
with the restrictions on the velocity 
\( u_1 = u_2 = u_3 = 0 \), 
it is known that the Kasner solution is a general one 
near the initial singularity \cite{kn:Pereseckii}. 
From the above, we can conclude that all these models 
are integrable. 

The facts that the BKL approximation represents a typical 
state of the metric in a very early state of the evolution 
has a very simple interpretation. Let for $F=F_1 \ll -1$ 
there be  some distribution of initial conditions (e.g. 
homogeneous). Approaching the initial 
singularity this distribution for $F=F_2$, where 
$F_2 < F_1$, transforms to the corresponding one 
concentrated in the neighborhood of the critical points. 
These critical points have the separatrices which move
towards the physical region of the phase space. 
During the motion along such a separatrix the space-time 
metric is described by the BKL approximation.

From the fact that 
\( F \rightarrow - \infty \) 
near the singularity, we can conclude the existence 
of a fundamental property of the system -- 
the property of concentration of the trajectories 
near the boundary $\Delta$. 

To finish with, we would like to make a certain 
suggestion more philosophical or methodological 
in  character. Among the Bianchi class A models,  
 the most general are BVIII and BIX.  
These models possess the highest dimension of 
space of structural constant.

So, the integrable subclass of the Bianchi models forms 
a set of zero measure in the space of all Bianchi class 
A models. This fact means that integrable cases are 
exceptional whereas those non-integrable ones are 
typical, as well as the Bianchi models which 
izotropize at infinite time in the full class of 
Bianchi models \cite{kn:Collins}. The establishing of the above 
fact for the Einstein equations in general is the challenge 
for the authors' future investigations.

\end{document}